\UseRawInputEncoding
\documentclass[aps, prl,reprint, amsmath,amssymb, superscriptaddress]{revtex4-2}

\usepackage{graphicx}
\usepackage{xcolor,color}

\usepackage[normalem]{ulem} 
\newcommand{\inserted}[1]{#1}

\begin{document}

\title{Time-resolved ARPES on cuprates:\\ Tracking the low-energy electrodynamics in the time domain}

\author{M.\,Zonno}
\email[]{mzonno@phas.ubc.ca}
\affiliation{Quantum Matter Institute, University of British Columbia, Vancouver, BC V6T 1Z4, Canada}
\affiliation{Department of Physics $\&$ Astronomy, University of British Columbia, Vancouver, BC V6T 1Z1, Canada}
\affiliation{Max Planck Institute for Solid State Research, Heisenbergstrasse 1, D-70569, Stuttgart, Germany}
\author{F.\,Boschini}
\email[]{fabio.boschini@inrs.ca}
\affiliation{Quantum Matter Institute, University of British Columbia, Vancouver, BC V6T 1Z4, Canada}
\affiliation{Centre \'{E}nergie Mat\'{e}riaux T\'{e}l\'{e}communications, Institut National de la Recherche Scientifique, Varennes, Qu\'{e}bec J3X 1S2, Canada}
\author{A.\,Damascelli}
\email[]{damascelli@physics.ubc.ca}
\affiliation{Quantum Matter Institute, University of British Columbia, Vancouver, BC V6T 1Z4, Canada}
\affiliation{Department of Physics $\&$ Astronomy, University of British Columbia, Vancouver, BC V6T 1Z1, Canada}

\begin{abstract}
\noindent The pursuit of a comprehensive understanding of the dynamical nature of intertwined orders in quantum matter has fueled the development of several new experimental techniques, including time- and angle-resolved photoemission spectroscopy (TR-ARPES).
In this regard, the study of copper-oxide high-temperature superconductors, prototypical quantum materials, has furthered both the technical advancement of the experimental technique, as well as the understanding of their correlated dynamical properties.
Here, we provide a brief historical overview of the TR-ARPES investigations of cuprates, and review what specific information can be accessed via this approach. We then present a detailed discussion of the transient evolution of the low-energy spectral function both along the gapless nodal direction and in the near-nodal superconducting gap region, as probed by TR-ARPES on Bi-based cuprates. 
\end{abstract}

\date{\today}

\maketitle

\section{Introduction}
The diverse and captivating properties of quantum materials emerge from the interplay between strong electron interactions and collective excitations \cite{keimerMoore2017review,basov2017NatMatReview}.
These properties are precariously balanced on the backdrop of multiple orders which compete and/or coexist in a dynamical fashion.
A variety of experimental techniques have been extended into the time domain via pump-probe stroboscopic approaches to enable the exploration of quantum phases of matter on their intrinsic timescales unattainable at equilibrium \cite{giannetti2016review}.
Among them, time- and angle-resolved photoemission (TR-ARPES)
is unique in its ability to directly access
the momentum-resolved electronic dynamics and interactions sub-picosecond timescales.
In the past two decades, TR-ARPES has witnessed significant advancements, offering new insights into the non-equilibrium properties of a variety of quantum materials \cite{schmitt2008Science,rohwer2011Nature,gierz2013NatMat,gerber2017Science,sobota2012PRL,wang2013Science,reimann2018subcycle,na2019Science,smallwood2016review,kemper2017review,kemper2018PRX}.
Copper-oxide high-temperature superconductors (HTSCs) are one set of quantum materials that motivated further experimental and theoretical advancements in TR-ARPES, resulting in unprecedented investigations of the dynamical properties of correlated materials \cite{perfetti2007,smallwood2012,rameau2016NatComm,parham2017prx,boschini2018}.
Cuprates are a prototypical example of a strongly correlated system: their phase diagram hosts a multitude of quantum phases whose origin and interplay are still under debate, such as (but not limited to) unconventional superconductivity, Mott insulating behavior, pseudogap phenomenon, charge order, and band-structure renormalization due to electron-boson coupling (the so called \textit{kink}) \cite{damascelli2003Review,davis2013PNAS,fradkin2015colloquium,comin2016Review,DopingMott_RMP2006}.
Here we offer a brief overview of the TR-ARPES research on cuprates over the past 15 years, as well as a comprehensive discussion of the transient evolution of the low-energy one-electron removal spectral function in Bi-based cuprates based on some of our recent works \cite{boschini2018,boschini2020npjQuantuMat,zonno2021PRB}, and new experimental data. 

\section{TR-ARPES on cuprates: \\a historical overview}
The undoped parent compounds of cuprates are characterized by a Mott antiferromagnetic insulating state driven by strong electron interactions \cite{keimerMoore2017review,damascelli2003Review,DopingMott_RMP2006}.
By removing or adding electrons to the CuO$_2$ plane, novel phases of matter emerge, including superconductivity, pseudogap, and charge order \cite{damascelli2003Review,hufner2008ReviewTwoGaps,comin2016Review,boschini2021NatComm,DopingMott_RMP2006}. The unconventional superconducting phase is defined by a $d$-wave order parameter. A gapless node is present along the Brillouin zone diagonal or, equivalently, at 45$^o$ with respect to the Cu-O bond in the CuO$_2$ plane (the so-called \textit{nodal} direction), while the maximum amplitude of the superconducting gap occurs at the edge of the Brillouin zone, the so-called \textit{antinodal} region \cite{ding1996PRBgap}. 

\begin{figure}[t]
\centering
\includegraphics[scale=1]{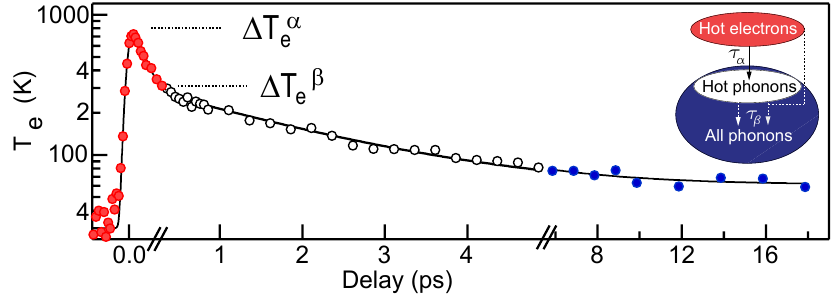}
\caption[FigR1]{Logarithmic plot of the transient evolution of the electronic temperature, measured for a 1.55\,eV pump with fluence of $\sim$100\,$\mu$J/cm$^2$. Following the optical excitation, two different relaxation regimes are observed: a fast exponential drop within $\tau_{\alpha}$=110\,fs, and a more gentle subsequent downturn with $\tau_{\beta}$=2\,ps as anharmonic decay and scattering from cold phonons occur. The inset sketches the energy transfer from hot-electrons to the phononic bath. Adapted from Ref.\,\onlinecite{perfetti2007}. }
\label{FigR1}
\end{figure}

Since about the year 2000, TR-ARPES has been employed to track the ultrafast relaxation processes of quasiparticles (QPs) along the nodal direction of Bi$_2$Sr$_2$CaCu$_2$O$_{8+\delta}$ (Bi2212). Specifically, by tracking the temporal evolution of the electronic temperature and applying a phenomenological three-temperature model (Fig.\,\ref{FigR1}), Perfetti et al. \cite{perfetti2007} showed that the photoexcited electronic population converges to a pseudo-equilibrium Fermi-Dirac distribution on a 50\,fs time scale, as a consequence of the strong electron-electron interactions present in the cuprates. Concurrently, energy is transferred from electrons to strongly-coupled phonons within 100\,fs; the photoexcited electrons then gradually relax back to equilibrium on a timescale of a few picoseconds via coupling to the whole phononic bath.
These results have not only extended previous all-optical studies into momentum space \cite{giannetti2016review}, but have laid the foundation for the forthcoming TR-ARPES studies of cuprates.
In the years following this initial study, other investigations of low-energy nodal excitations in Bi-based cuprates have shown that: (i) the transient QP signal bears a relation to the superconducting, pairing, and pseudogap onset temperatures \cite{zhang2013prb};  (ii) the nodal electron dynamics are related to superconducting correlations even above T$_c$ \cite{piovera2015prb}; and (iii) the QP population lifetime does not share a direct relation to the single-particle lifetime \cite{yang2015inequivalence}.

Following this, an effort was made to perform similar studies away from the gapless nodal direction and investigate how the $d$-wave superconducting gap may affect QP dynamics within (and slightly beyond) the Fermi arc region. Although the precise momentum dependence of the QP relaxation has been the subject of intense debate, all studies have reported a $\approx$\,2-5\,ps recombination time of the photoemission intensity within the Fermi arc region in the superconducting state. The recombination phenomenology has been discussed in terms of bimolecular recombination of QPs into Cooper pairs, boson-bottleneck effects, or phase-space scattering restrictions \cite{cortes2011,smallwood2012,smallwood2015prbR,zhang2016sciRep,konstantinova2018SciAdv,cilento2014NatComm}.
Finally, with the successful development of high-harmonic-based TR-ARPES systems, Dakovski et al. \cite{dakovski2015quasiparticle} performed the first survey of QP dynamics at the antinodal region of Bi2212 in 2015. This study reported a non-monotonic temporal evolution of the photoemission intensity following the pump excitation, remarkably different from what was observed within the near-nodal region \cite{dakovski2015quasiparticle}.

\begin{figure}[b]
\centering
\includegraphics[scale=1]{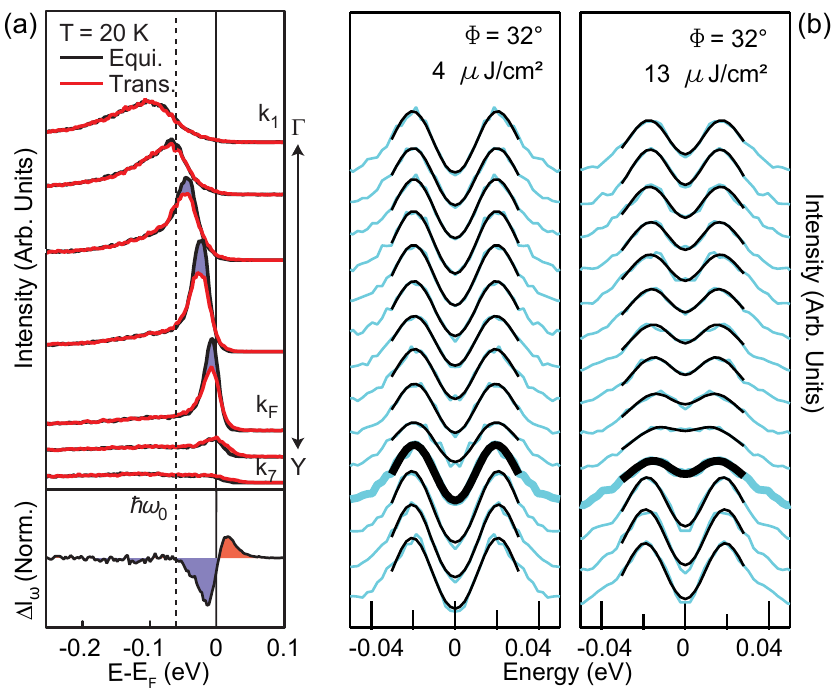}
\caption[FigR2]{(a) Top: nodal EDCs for equlibrium (black) and transient state (red); bottom: difference between the transient and equilibrium EDCs, integrated through a momentum range centred around k$_\mathrm{F}$, highlighting the spectral weight loss (blue) and gain (red). Adapted from Ref.\,\onlinecite{graf2011NatPhys}.
(b) Transient evolution of the near-nodal superconducting gap within 8\,ps after the 1.55\,eV pump excitation (gap magnitude at equilibrium $\approx$15 meV), for two different pump fluences. Symmetrized EDCs are described by a phenomenological model which includes a single scattering term and the superconducting gap amplitude; the bold curves correspond to the zero pump-probe delay. Adapted from Ref.\,\onlinecite{smallwood2012}.}
\label{FigR2}
\end{figure}

The inspection of the transient evolution of the momentum- and/or energy-resolved photoemission intensity has offered significant insights into the QPs dynamics in cuprates. However, the unique advantage of ARPES is its capability to access the one-electron removal spectral function in momentum space, providing direct insights into underlying electron interactions embedded in the many-body self-energy \cite{damascelli2003Review}. 
Initial studies with ultrafast methods reported a transient melting of the nodal QP spectral weight across T$_c$ \cite{graf2011NatPhys} [Fig.\,\ref{FigR2}(a)]. This has been recently discussed in terms of a direct manifestation of the energy and temperature dependence of the single-particle lifetime within the Fermi liquid formalism \cite{zonno2021PRB}.

A major milestone in the study of the physics of cuprates via TR-ARPES came in 2012, when Smallwood et al. \cite{smallwood2012} demonstrated, for the first time, the ability to track the evolution of the near-nodal superconducting gap in Bi2212 with sub-picosecond temporal resolution [see Fig.\,\ref{FigR2}(b)]. This seminal work 
laid the basis for the TR-ARPES exploration of the superconducting gap in high-temperature superconductors and was soon followed by several other studies \cite{smallwood2014prb,parham2017prx,zhang2017prb,boschini2018}. While early investigations have reported a photoinduced gap closure in the near-nodal region \cite{smallwood2012,smallwood2014prb}, subsequent studies revealed that the gap amplitude is minimally affected upon near-IR excitation and, instead, the gap fills in as a consequence of the loss of coherence of the superconducting condensate \cite{parham2017prx,zhang2017prb,boschini2018}. 

\begin{figure}[t]
\centering
\includegraphics[scale=1]{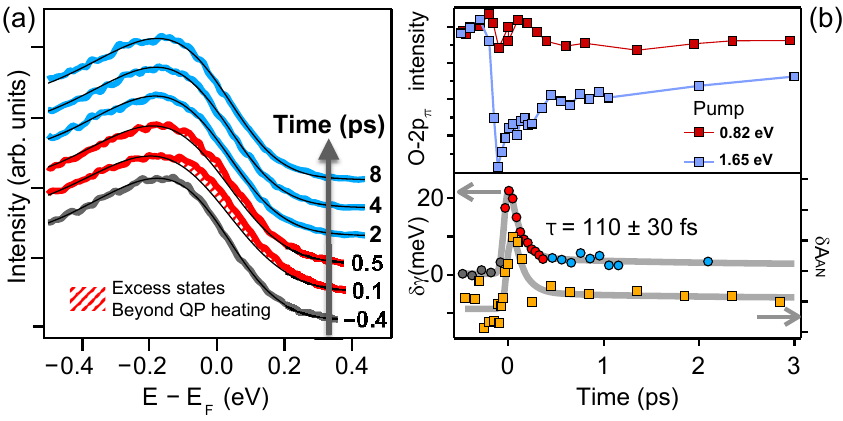}
\caption[FigR3]{(a) Nonequilibrium antinodal EDCs measured at different time delays after a 1.65 eV pump excitation. The dashed areas show the excess signal, with respect to a simple effective heating, related to the transient increase of states at the Fermi level ($\delta$A$_\text{AN}$). 
(b) Top: Dynamics of the photoemission intensity at a binding energy of 1.2 eV (O-2p$_\pi$ band) for the 0.82 eV (red squares) and 1.65 eV (blue squares) pump excitations. Bottom: comparison between the broadening of the O-2p$_\pi$ peak (circles, left axis) and antinodal increase of states $\delta$A$_\text{AN}$ (yellow squares, right axis). Adapted from Ref.\,\onlinecite{cilento2018SciAdv}.}
\label{FigR3}
\end{figure}

In contrast to the extensive efforts made to study the superconducting gap, the transient evolution of the pseudogap has yet to receive comparable attention. This is likely due to the limited momentum-space accessible to low-photon energy laser systems, which were preferred in the developmental years of the TR-ARPES technique for their easy implementation \cite{gauthier2020_JAP_6eVsystems}. 
To date, only two TR-ARPES studies have directly tracked the transient evolution of the pseudogap. This was done at the antiferromagnetic hot spot for the electron-doped cuprate Nd$_{\text{2-x}}$Ce$_{\text{x}}$CuO$_{\text{4}}$ \cite{boschini2020npjQuantuMat}, and at the antinode in the hole-doped bilayer material Bi2212 \cite{cilento2018SciAdv}.
In the latter case, the appearance of antinodal in-gap states [see Fig.\,\ref{FigR3}(a)] has been reported following a resonant transfer of electrons between Cu and O orbitals, as shown in Fig.\,\ref{FigR3}(b). This pioneering study of the antinodal dynamics of Bi2212 demonstrates the close relation between low- and high-energy scales, in support of a correlation-driven origin of the pseudogap \cite{cilento2018SciAdv}. 

\begin{figure}[b]
\centering
\includegraphics[scale=1]{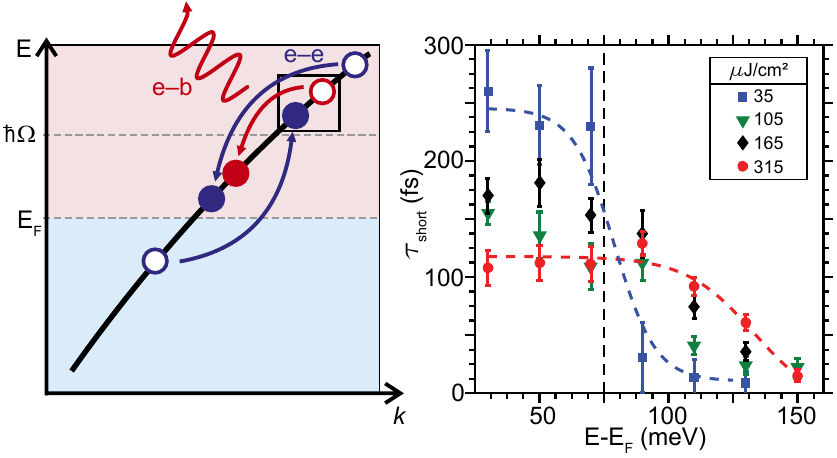}
\caption[FigR4]{Left: illustration of the dichotomy between electron-boson (e-b) and electron-electron (e-e) scattering contributions to the population dynamics. Right: experimental fast decay time scale of the photoemission intensity above the Fermi level as a function of the pump fluence. The bosonic mode energy at 75\,meV is marked by the dashed black line. Adapted from Ref.\,\onlinecite{rameau2016NatComm}.}
\label{FigR4}
\end{figure}

The journey of the TR-ARPES community in exploring the non-equilibrium properties of cuprates did not stop at the study of the QPs dynamics and quenching of the superconducting gap. In 2014 Rameau et al. \cite{rameau2014prb} and Zhang et al. \cite{zhang2014NatComm} separately investigated the transient modification of the band renormalization at approximately 70\,meV binding energy (known as the \textit{kink} \cite{lanzara2001_NatureKink}) in Bi-based cuprates. While the former \cite{rameau2014prb} reported a sub-100\,fs renormalization of the nodal effective mass which was explained to be the result of the dissolution of electronic correlations by the optical pump \cite{sentef2013PRX}, the latter \cite{zhang2014NatComm} uncovered a direct link between the quenching of the real-part of the electron self-energy at the kink energy and the melting of the superconductive condensate, thus suggesting a photoinduced reduction of the electron-boson coupling constant. The relation between the superconducting gap and the nodal kink softening has been more recently confirmed \cite{ishida2016SciRep}, and extended to the off-nodal region by analyzing the dip-hump structure \cite{miller2018prb}. 
Moreover, by exploiting the ability of TR-ARPES to access QP scattering processes in the unoccupied states, Rameau et al. \cite{rameau2016NatComm} reported an enhanced recombination rate for energies above the boson window ($\omega >$70\,meV, sub-30\,fs decay; Fig.\,\ref{FigR4}), suggesting the involvement of a selective electron-boson coupling with remarkable similarities to the kink phenomenon observed in the occupied states.

In addition to all the investigations listed above, several other alternative applications of TR-ARPES on cuprates have been demonstrated in recent years. These include: (i) tracking of the transient shift of the chemical potential along the nodal direction, and its relation to pump-induced modifications of the superconducting gap \cite{miller2015prb} and the particle-hole asymmetric pseudogap \cite{miller2017prl}; (ii) mapping of electronic gaps \cite{miller2015prbUnoccupied} and the upper Hubbard band \cite{yang2017prb} in the unoccupied states; and (iii) extraction of the momentum-dependent deformation potential via photoinduced coherent phonons \cite{yang2019prl}.

\begin{figure*}[t]
\centering
\includegraphics[scale=1]{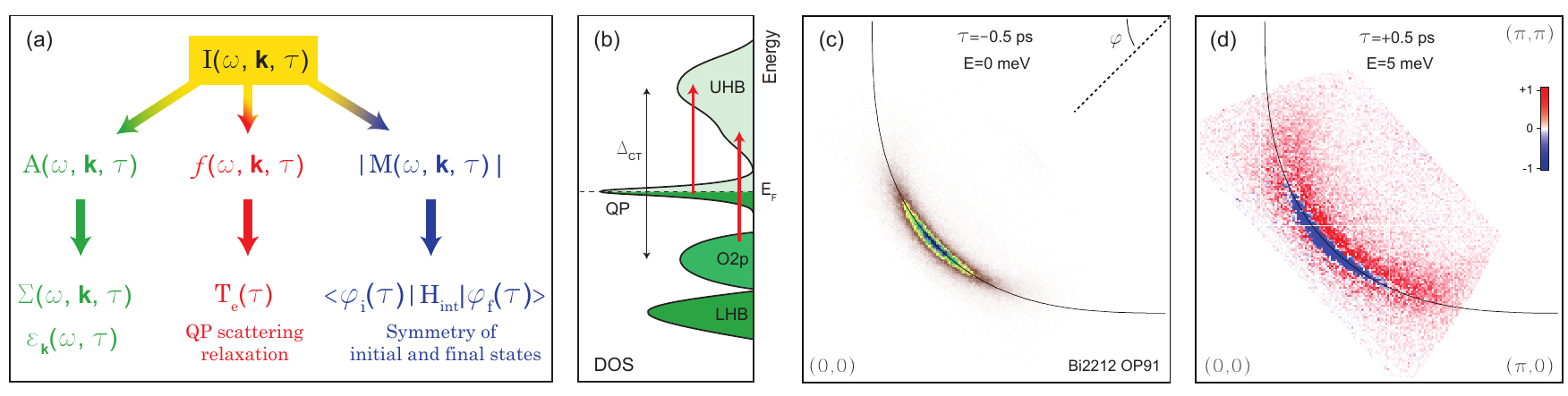}
\caption[Fig2]{(a) Sketch of the different contributions to the transient photoemission intensity $\text{I}(\omega,\textbf{k},\tau)$. 
(b) Schematic of the density of states of the doped cuprates (adapted from Ref.\,\onlinecite{baldini2020PNAS}). Near-IR excitations promote carriers from the O-2p and low-energy electronic states (where coherent QPs are present) into the unoccupied upper Hubbard band (UHB) \cite{cilento2018SciAdv}. The lower Hubbard band (LHB) and the charge-transfer gap ($\Delta_{\text{CT}}$) are also indicated.
(c) Fermi surface mapping of optimally-doped Bi2212 (T$_c$=91\,K, OP91) at $\omega$=0\,meV (6\,meV integration window), before the 1.55\,eV pump excitation. 
(d) Differential iso-energy contour [$\text{I}(\tau=$ 0.5\,ps) $-$ $\text{I}(\tau=-$0.5\,ps)] at 5\,meV above the Fermi level, 20\,meV integration window. The sample is aligned along the $\Gamma$–Y direction and the solid black line represents a tight-binding constant energy contour at $\omega$=0 \cite{kaminski2005mPRB_TB}. The dashed black line in (c) defines the angle $\varphi$ of a radial cut.
}
\label{Fig2}
\end{figure*}

As a final consideration to conclude this brief overview of the TR-ARPES investigations of cuprates, one may notice that most of the aforementioned works have been performed using a near-IR pump excitation and low-energy ($\sim$\,6\,eV) probe photons \cite{gauthier2020_JAP_6eVsystems}, which are intrinsically limited to the study of the near-nodal momentum region. Even though high-harmonics-based TR-ARPES systems have been available for more than a decade, only very recently has the combination of high-stability, high-repetition rate, and sub-50\,meV energy resolution been achieved \cite{sie2019NatComm,mills2019RSI,lee2020RSI,puppin2019RSI}, thus paving the way towards fascinating studies of low-energy electrodynamics over the whole Brillouin zone of cuprates, as well as other quantum materials.
As for the optical excitation, despite its easy accessibility, the commonly-used $\sim$\,1.5\,eV pump energy bears some critical limitations, as it redistributes carriers around the entire momentum space, without coupling to any specific collective mode (whose energy scales are commonly sub-100\,meV) \cite{basov2017NatMatReview,giannetti2016review,mankowsky2016Review}. Following the path already set by all-optical studies, where tailored mid-infrared/THz excitations have been used to resonantly excite specific collective modes and drive new phases of matter with no equilibrium counterparts \cite{rini2007Nature,fausti2011Science,mitrano2016Nature,giusti2019PRL,mankowsky2016Review}, next-generation TR-ARPES systems will offer long-wavelength pump capabilities, enabling new insights into the light-induced manipulation of the electronic properties of quantum materials with the addition of direct momentum resolution. The potential of these types of experiments has recently been demonstrated in several pioneering works \cite{wang2013Science,reimann2018subcycle,aeschlimann2021GierzArXiv}.

\section{TR-ARPES intensity}
In the previous section, we have discussed the types of valuable dynamical information that TR-ARPES has been capable of extracting in cuprates so far. However, one should note that the ARPES intensity (and consequently its extension to the time domain) is  the manifestation of a complex interplay between different contributions, which often make its interpretation challenging. In particular, the equilibrium ARPES intensity for fixed energy $\omega$ and momentum $\textbf{k}$ can be defined via Fermi's Golden rule as $\text{I}(\omega,\textbf{k})=A(\omega,\textbf{k}) \cdot |M(\omega,\textbf{k})|^2 \cdot f(\omega)$,
where $A(\omega,\textbf{k})$ is the one-electron removal spectral function, $|M(\omega,\textbf{k})|^2$ the photoemission matrix element, and $f(\omega)$ the Fermi-Dirac distribution function \cite{damascelli2003Review}. When considering optically-induced electrodynamics, one may approximate the TR-ARPES signal by extending the equilibrium formalism into the time domain $\tau$ as follow:
\begin{equation}
    \label{EQ1}
    \text{I}(\omega,\textbf{k},\tau) \simeq A(\omega,\textbf{k},\tau) \cdot |M(\omega,\textbf{k},\tau)|^2 \cdot f(\omega,\textbf{k},\tau).
\end{equation}
Note that this approximation treats the photoemission intensity at each time-delay as a thermally-broadened equilibrium ARPES signal, and does not describe the effects of the pump pulse's electric field, as well as other highly non-thermal processes \cite{sentef2013PRX,randi2017PRB,kemper2017review,kemper2018PRX}. However, given their intrinsic $\sim$100\,fs thermalization time \cite{perfetti2007}, Eq.\,\ref{EQ1} offers a reasonable description of the TR-ARPES signal of cuprates for $\tau >$100-200\,fs.

Figure\,\ref{Fig2}(a) illustrates the different contributions to the transient photoemission intensity of Eq.\,\ref{EQ1}. While $A(\omega,\textbf{k},\tau)$ encompasses the transient modifications of the many-body self-energy $\Sigma(\omega,\textbf{k},\tau)$ \cite{smallwood2012,zhang2014NatComm,parham2017prx,boschini2018,schmitt2008Science,rohwer2011Nature,zonno2021PRB}, and of the electronic band dispersion $\epsilon_k(\omega,\tau)$ \cite{yang2019prl,gerber2017Science,sobota2014PRLphonons,hein2020NatCommFTarpes,wang2013Science,nicholson2018Science}, the transient electronic distribution $f(\omega,\textbf{k},\tau)$ follows intrinsic QP scattering and relaxation processes \cite{perfetti2007,graf2011NatPhys,sobota2012PRL,yang2015inequivalence,na2019Science,na2020PRB,zhang2016sciRep, dakovski2015quasiparticle,gierz2013NatMat,monney2016TiSe2PRB,kuroda2016PRL_midIR_TI,rohde2018PRL_grapheneSub50fs,chen2020PNAS_InSe}.
Finally, the transient modification of the photoemission matrix element $M(\omega,\textbf{k},\tau)$ has recently been acknowledged theoretically \cite{freericks2016matrixElement}, and demonstrated experimentally \cite{boschini2020NJP}. $M(\omega,\textbf{k},\tau)$ connects the initial ($\varphi_i$) and final ($\varphi_f$) state of the photoemitted electron \cite{day2019Chinnok}, and its temporal evolution may be significant in multi-orbital systems, implying dynamical changes of the orbital mixing of the probed states. In the specific case of cuprates, TR-ARPES mainly accesses a single band with a well-defined orbital character (\textit{i.e.} d$_{x^2-{y^2}}$), hence transient modifications of $M$ may be neglected in a first order approximation.
Therefore, the transient photoemission intensity of cuprates reflects primarily changes in both $f(\omega,\textbf{k},\tau)$ and $A(\omega,\textbf{k},\tau)$, which encode the pump-induced redistribution of carriers and changes in the electron self-energy, respectively \footnote{Note that a transient modification of the self-energy is a common feature of all TR-ARPES measurements, not only cuprates: as a paradigmatic example, pump-induced photoholes in the occupied manifold necessarily lead to a reduction of the single-particle lifetime, thus broadening the spectral features.}.
This implies that $\text{I}(\omega,\textbf{k},\tau)$ cannot be generally considered a direct and straightforward measure of the intrinsic transient properties of a material, such as QP dynamics. Instead, extensive experimental efforts and data analysis efforts are needed to disentangle the distinctive contributions to the transient photoemission intensity from either the electronic distribution or the spectral function (and more generally the matrix elements).

In the following, we present a detailed analysis of the pump-induced evolution of the spectral function and many-body self-energy in single- and double-layer Bi-based cuprates within the Fermi arc region. All TR-ARPES investigations presented here were conducted by employing low-fluence (sub-90 $\mu$J/cm$^2$) 800\,nm (1.55\,eV) pump excitations. The diagram in  Fig.\,\ref{Fig2}(b) shows that a near-IR optical excitation redistributes carriers from the electronic states close to the Fermi level and O-2p bands, as recently demonstrated by Cilento et al. \cite{cilento2018SciAdv}. As a direct consequence of electron interactions and scattering from high-energy bosonic modes \cite{dalConte2012Science}, highly non-thermal electrons thermalize in the proximity of the Fermi level (below the 70\,meV bosonic window) within $\sim$100\,fs \cite{perfetti2007,rameau2016NatComm}. 
To illustrate how the interplay between the different contributions in Eq.\,\ref{EQ1} may affect the TR-ARPES signal of Bi-based cuprates, we present in Fig.\,\ref{Fig2}(c),(d) the Fermi surface mapping of optimally-doped Bi2212 (T$_c$=91\,K, OP91) before the pump excitation, and the differential iso-energy contour mapping at 5\,meV above the Fermi level 0.5\,ps after excitation.
Along the nodal direction ($\varphi=45^o$), we observe both a suppression and an increase of intensity [blue and red in Fig.\,\ref{Fig2}(d), respectively] as a direct consequence of modifications of the electronic distribution (due to higher electronic temperature), as well as of the self-energy (as a result of suppression of the coherent spectral weight and/or spectral broadening). Moreover, beyond the near-nodal region, a clear increase of the photoemission intensity is present, which can be attributed to a transient filling of the superconducting gap and to thermal broadening. 
In the next sections, we will discuss both of these aspects, starting from the analysis of the gapless nodal spectral features and then moving on to the study of the near-nodal dynamics.

\begin{figure*}[]
\centering
\includegraphics[scale=1.03]{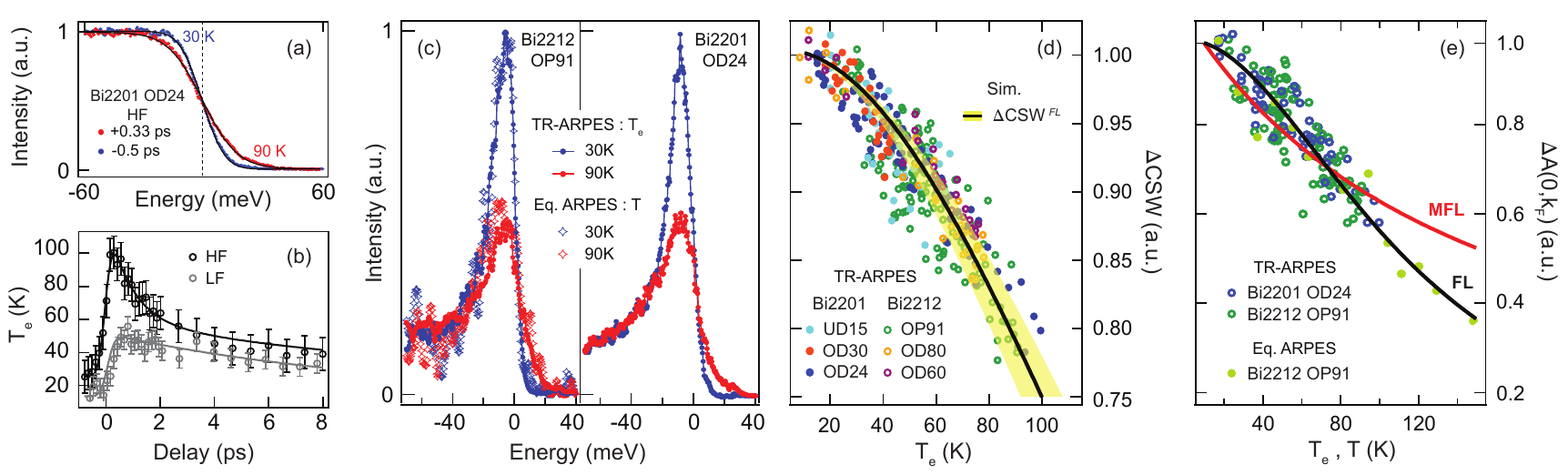}
\caption[Fig3]{(a) Momentum-integrated nodal energy distribution curves (EDCs) before (-0.5\,ps) and after (+0.33\,ps) the pump excitation for Bi2201-OD24. Fermi-Dirac distribution fits are shown as solid black lines. (b) Transient electronic temperature T$_\mathrm{e}$ obtained by fitting the Fermi edge width of momentum-integrated nodal EDCs of Bi2201-OD24, for two different pump fluences ($12\,\mu$J/cm$^2$, LF; $90\,\mu$J/cm$^2$, HF). The solid lines represent (single-) double-exponential decay fit to (LF) HF data. (c) EDCs at the Fermi momentum $\mathbf{k}_\mathrm{F}$ along the nodal direction of Bi2212-OP91 (left) and Bi2201-OD24 (right) at \inserted{T$_\mathrm{e}$=30\,K (LF) and T$_\mathrm{e}$=90\,K (HF)}, as measured by TR-ARPES with $h\nu$=$6.2$\,eV probe photon energy. \inserted{The sample base temperature at negative pump-probe delays has been estimated via Fermi-Dirac distribution fit to be 25\,K (45\,K) at LF (HF) for Bi2212-OP91 and 15\,K (25\,K) at LF (HF) for Bi2201-OD24.} Equilibrium ARPES ($h\nu$=$27$\,eV) data are also shown for Bi2212-OP91 \inserted{at the two sample temperatures.} All EDCs have been deconvoluted from the energy resolution via the Lucy-Richardson algorithm \cite{razzoli2013PRL,yang2008NatureDeconv} (TR-ARPES energy resolution is 11\,meV and 18\,meV for left and right panels, respectively; equilibrium ARPES energy resolution is 5.3\,meV). (d) Comparison between the experimental relative variation of the nodal coherent spectral weight $\Delta$CSW for different Bi-based compounds (colored data points, integration window $[-0.08, 0.08]\,$eV, normalized to the 10\,K value) and the simulated $\Delta$CSW$^{FL}$ within the FL model obtained via Eq.\,\ref{eq1FL}. The solid black line is computed for $\alpha$=0.7, $\beta$=20.75, $\Gamma_{\text{imp}}$=0.02, while the yellow shading accounts for the range of parameters present in the different compounds. (e) Normalized temperature evolution of the spectral function $\Delta A(0,\mathbf{k}_\mathrm{F})$, for Bi2201-OD24 and Bi2212-OP91. The integration range in momentum and energy is $0.005\,$\AA$^{-1}$ and $6\,$meV, respectively. The solid lines are best fits to the experimental data using Eq.\,\ref{eq2FL} for the two different models with $\Gamma_{\text{imp}}$=0.02: FL (black, $\beta$=20.35), MFL (red, $\lambda$=0.96).}
\label{Fig3} 
\end{figure*}

\section{Nodal Coherent Spectral Weight \\ and electrodynamics}
We begin the exploration of the TR-ARPES signal in Bi-based cuprates from the nodal direction, which is not influenced by the superconducting gap and pseudogap phenomenon \cite{vishik2012PNAS}.
Specifically, we focus our attention on the origin of the coherent spectral weight (CSW). The emergence of coherent QPs has been identified as a signature of the superconducting state and linked to its characteristic onset temperature T$_c$ \cite{sawatzky1989testing,shen1999novel}, leaving the question of if and how such coherent excitations are present also in the normal state. Equilibrium and out-of-equilibrium ARPES experiments on bilayer Bi2212 have reported a suppression of the CSW of both the nodal and antinodal QPs as a function of the increasing temperature, exhibiting an abrupt modification around T$_c$. This suggests a direct relation to the coherence of the superconducting condensate \cite{feng2000signature,ding2001coherent,graf2011NatPhys}, although a theoretical work has questioned a direct link between CSW and superconductivity \cite{zheng2017coherent}. 

Towards a conclusive description of the evolution of the CSW across the superconducting-to-normal-state phase transition in cuprates, we recently carried out a comprehensive temperature-dependent study of the nodal CSW for various doping levels of both single- and bi-layer Bi-based compounds \cite{zonno2021PRB}. In this work we employed TR-ARPES as an alternative tool to conventional equilibrium ARPES to perform a detailed temperature-dependent study of the nodal spectral features, by finely tuning the electronic temperature T$_e$ via optical pumping. This time-resolved approach relies on the fact that a quasi-equilibrium state is reached by the system after a near-IR excitation. As previously discussed, pump-induced non-thermal electrons and charge-redistribution thermalize within 100\,fs, allowing for an effective electronic temperature to be defined at each pump-probe delay $>$\,100-200\,fs \cite{perfetti2007,graf2011NatPhys,boschini2020npjQuantuMat,cilento2018SciAdv,giannetti2016review}. 

The determination of T$_e (\tau)$ is achieved by fitting the Fermi edge of the momentum-integrated energy distribution curves (EDCs) along the nodal direction for each delay, as illustrated in Fig.\,\ref{Fig3}(a) for overdoped Bi$_2$Sr$_2$CuO$_{6+\delta}$ (Bi2201, T$_c$=$24\,$K, OD24). Under the approximations of a linear dispersion, momentum-independent self-energy, and constant matrix-elements, the momentum-integrated signal is proportional to the electronic Fermi-Dirac distribution \cite{reber2012TDOS}, and once the experimental resolution has been accounted for, its width provides a measure of the effective electronic temperature T$_e$. 
Following this procedure, the full time dependence of T$_e (\tau)$ can be retrieved [Fig.\,\ref{Fig3}(b)] and correspondingly linked to the time-dependent changes in the nodal spectral features. This facilitates a temperature-dependent study, which is minimally affected by degradation of the cleaved surface and drift of experimental conditions during the macroscopic temperature cycling of cryogenic manipulators. Figure\,\ref{Fig3}(c) shows the nodal EDCs at the Fermi momentum $\mathbf{k}_\mathrm{F}$ of optimally doped Bi2212 (T$_c$=$91\,$K, OP91) and overdoped Bi2201 (T$_c$=$24\,$K, OD24) as acquired by TR-ARPES; a clear suppression of CSW close to the Fermi energy (E$_\mathrm{F}$) is observed in both compounds as the transient T$_e (\tau)$ increases from 30\,K to 90\,K. In order to exclude any possible non-thermal contributions to the reported quenching of spectral weight, we performed complementary T-dependent equilibrium ARPES measurements on Bi2212-OP91 and plot the nodal EDCs in Fig.\,\ref{Fig3}(c), left panel. 
\inserted{Note that although the lattice and electronic temperatures may differ within the first 1-2\,ps, the excellent agreement between equilibrium and time-resolved ARPES establishes the electronic temperature in the system as the main parameter defining the suppression of CSW, and validates our use of optical-pumping in TR-ARPES to control the electronic temperature of the system.} 

We tracked the temperature evolution of the nodal CSW at $\mathbf{k}_\mathrm{F}$, CSW($\mathbf{k}_\mathrm{F}$,\,T)=$\int_{-\infty}^{\infty}A(\omega,\mathbf{k}_\mathrm{F},\text{T}) \, d \omega$, by estimating the relative variation $\Delta$CSW via the difference of integrated area under the symmetrized EDCs at $\mathbf{k}_\mathrm{F}$ at low and high T$_\mathrm{e}$ (\textit{i.e.} before and after the pump excitation; integration window $[-0.08, 0.08]\,$eV, normalized to the 10\,K value) \cite{zonno2021PRB}.
We recall that for a particle-hole symmetric spectral function (which has been experimentally verified at $\mathbf{k}_\mathrm{F}$ in the near-nodal region of Bi-based cuprates \cite{matsui2003PRL_ph_symm}), the symmetrization procedure removes the contribution of the Fermi-Dirac distribution \cite{norman1998Nature}.
Figure\,\ref{Fig3}(d) displays the resulting $\Delta$CSW for different doping levels of both single- and bi-layer Bi-cuprates (full and open circles, respectively). These results uncover two important aspects: (i) despite the variety of dopings and T$_c$, the suppression of CSW is ubiquitous across the studied compounds, with a quenching as large as $\sim$\,25$\%$ at 100 K; (ii) the decrease in nodal CSW extends past T$_c$ without exhibiting any sudden change across the superconducting-to-normal state phase transition, and thus excludes a direct relation between the evolution of the nodal CSW and the superconducting phase.

The progressive quenching of the nodal spectral weight can be interpreted in terms of the intrinsic $\omega$- and T-dependence of the self-energy within the Fermi liquid (FL) theory. The FL self-energy can be written as \cite{damascelli2003Review,pines1966theory}:
\begin{equation} \label{eq1FL}
\Sigma _{\text{FL}} (\omega, \text{T}_e)  = -\alpha \omega -i [ \Gamma_{\text{imp}} +\beta\, (\omega^2 + \pi^2 k_B^2 \text{T}_e^2)] \, .
\end{equation}
Here, $\alpha$, $\beta$, and $\Gamma_{\text{imp}}$ are positive parameters, with the latter describing a scattering rate term related to static impurities and is thus energy and temperature independent \cite{abrahams2000angle}. 
Using Eq.\,\ref{eq1FL} we can calculate the T-dependence of the nodal spectral function and the expected variation of spectral weight, $\Delta$CSW$^{FL}$. In order to do so, the FL parameters $\alpha$ and $\beta$ were first estimated via a global fit of energy- and momentum-distribution curves via Eq.\,\ref{eq1FL} for different temperatures of Bi2212-OP91 and Bi2201-OD24 \cite{zonno2021PRB}. The best fits were achieved for ($\alpha$=0.8, $\beta$=20) and $(\alpha$=0.6 and $\beta$=21.5), respectively, which are consistent with what has been reported in previous ARPES studies \cite{kordyuk2005bare, zhou2003universal, zhang2014NatComm}. Using this range of parameters, we calculated $\Delta$CSW$^{FL}$ at $\mathbf{k}_\mathrm{F}$ and compared it to the experimental data in Fig.\,\ref{Fig3}(d) (solid black line and yellow shadow; integration window $[-0.08, 0.08]\,$eV, normalized to the 10\,K value). The FL-simulated suppression of nodal spectral weight agrees remarkably well with the experimentally observed evolution over the entire temperature range explored. 

To further support our interpretation of the nodal CSW in terms of the FL phenomenology, we analyze the spectral weight at $\omega$=0 and $\mathbf{k}_\mathrm{F}$, which allows for a simpler and more direct comparison to existing models such as FL and marginal FL (MFL) \cite{varma1989phenomenology}. The spectral function $A(0,\mathbf{k}_\mathrm{F},\text{T}_e)$ can be expressed as:
\begin{equation} \label{eq2FL}
A(0,\mathbf{k}_\mathrm{F},\text{T}_e)=
\begin{cases}
\frac{1}{\pi}\, \frac{1}{\Gamma_{\text{imp}}+\beta(\pi k_B \text{T}_e)^2} \, & \text{for FL}\\
\frac{1}{\pi}\, \frac{1}{\Gamma_{\text{imp}}+ \lambda(\frac{\pi}{2} k_B \text{T}_e)} \, & \text{for MFL},\\
\end{cases}
\end{equation}
where $\beta$ and $\lambda$ fully define the T-dependence in the two different models, and $\Gamma_{\text{imp}}$ is a temperature and energy independent scattering rate, as in Eq.\,\ref{eq1FL}. In Fig.\,\ref{Fig3}(e) we present the relative variation of the $\omega$=0 spectral function $\Delta A(0,\mathbf{k}_\mathrm{F})$, for Bi2201-OD24 and Bi2212-OP91, along with the best fit to the data for the two different models of Eq.\,\ref{eq2FL} (solid lines, normalized to the 10\,K value). While the MFL description fails to reproduce the experimental T-evolution, FL provides an excellent agreement up to 150\,K, returning a value of $\beta$=20.3 that is fully consistent with what was obtained via EDC-MDC global fit and used for the simulation in Fig\,\ref{Fig3}(d) \cite{zonno2021PRB}.

These findings demonstrate that the observed T-dependent evolution of the nodal coherent spectral weight is ubiquitous across the Bi-based compounds and bears no relation to the superconducting condensate. Instead, the suppression of CSW at $\mathbf{k}_\mathrm{F}$ naturally stems from the temperature and energy dependence of the complex electron self-energy within the Fermi liquid framework, thus indicating that nodal QPs retain the FL character above and below T$_c$.

\begin{figure}
\centering
\includegraphics[scale=1.05]{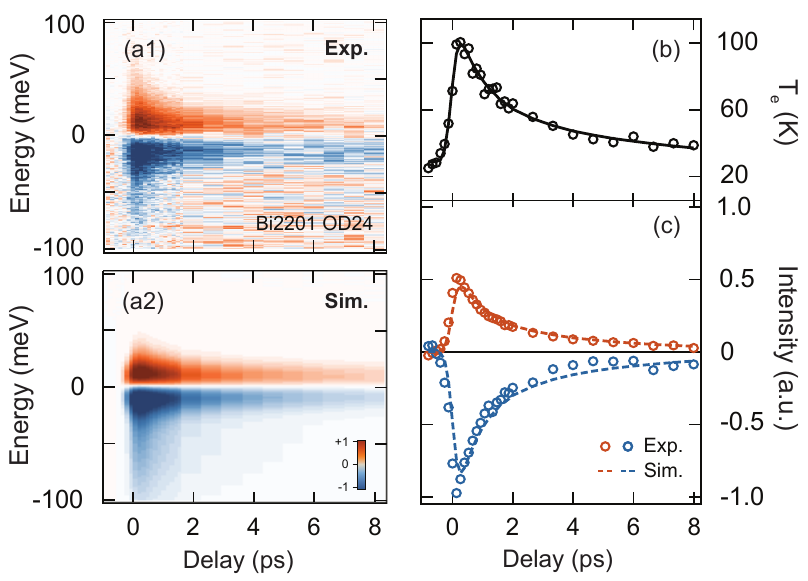}
\caption[Fig2]{(a) Experimental (a1) and simulated (a2) differential momentum-integrated EDCs along the nodal direction of Bi2201-OD24. The simulated map has been generated via Eq.\,\ref{eq1FL} using T$_e (\tau)$ shown in panel (b) and $\alpha$, $\beta$, and $\Gamma_{\text{imp}}$ parameters as in Fig.\ref{Fig3}. (b) Transient electronic temperature. The solid line is a phenomenological double exponential fit, with decay times $\sim$1.3\,ps and $\sim$6.4\,ps. (c) Population [orange; integration window (5, 55)\,meV] and depletion [blue; integration window (-55, -5)\,meV] dynamics, extracted from the differential maps in panel (a). Similar decay times characterize both signals, namely $\sim$0.85\,ps and $\sim$4.5\,ps as extracted via a phenomenological double exponential fit.}
\label{Fig4}
\end{figure}

\begin{figure*}
\centering
\includegraphics[scale=1.02]{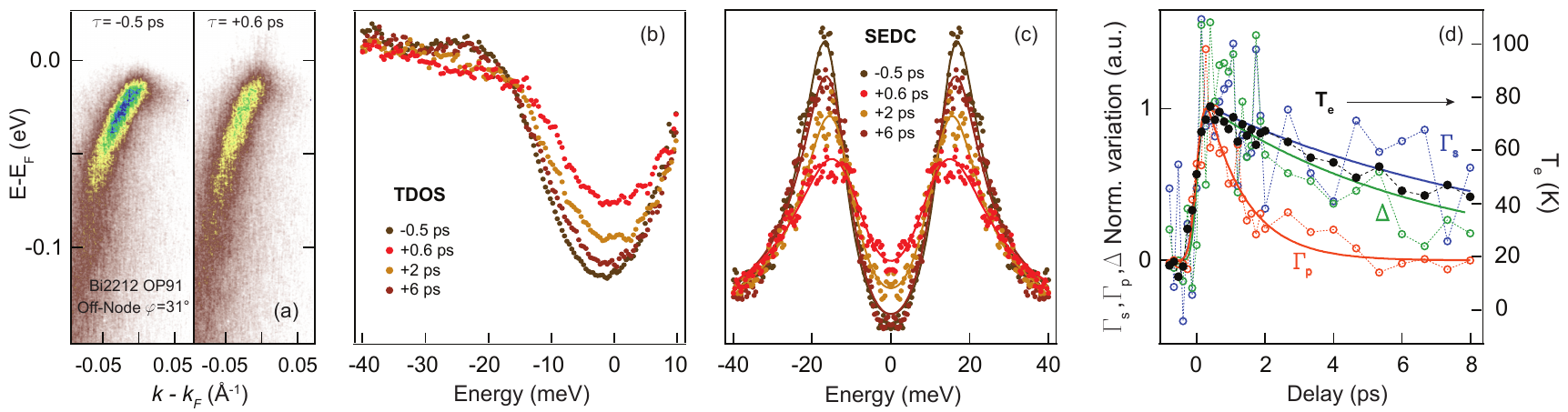}
\caption[Fig5]{(a) TR-ARPES spectra of Bi2212-OP91 acquired for $\varphi$=31$^o$ [where $\varphi$ defines the angle from the ($\pi$,0)-($\pi$,$\pi$) direction; see Fig.\,\ref{Fig2}(a)], before and after the pump excitation. (b) Transient tomographic density of states (TDOS) for selected pump-probe delays. (c) Symmetrized EDCs at $\mathbf{k}_\mathrm{F}$ (full circles) for the same delays as in panel (b), along with the corresponding fits obtained by using Eq.\,\ref{EQ2} (solid lines). (d) Transient normalized variation of the parameters $\Gamma_s$, $\Delta$, and $\Gamma_p$ (left axis), as obtained via the fit of symmetrized EDCs to Eq.\,\ref{EQ2}. The filled black circles represent the temporal evolution of the electronic temperature extracted along the nodal direction (right axis).}
\label{Fig5}
\end{figure*}

In the discussion above, we addressed the temperature-dependence of the nodal spectral function by symmetrizing (following Ref. \cite{norman1998Nature}) the EDCs to remove the contribution of the Fermi-Dirac distribution $f(\omega,\tau)$. However, the analysis of TR-ARPES data relies more conventionally on tracking the transient evolution of the unsymmetrized momentum-integrated nodal photoemission intensity $\int_k \text{I}(\omega,\textbf{k},\tau)dk$, which, as depicted in Fig.\,\ref{Fig2}(a), depends on both $A(\omega,\tau)$ and $f(\omega,\tau)$. In this regard, one may wonder whether the transient evolution of the nodal $\int_k \text{I}(\omega,\textbf{k},\tau)dk$ upon near-IR pumping can be fully rationalized within the FL framework by taking into account the dependence of both $A(\omega,\tau)$ and $f(\omega,\tau)$ on the effective electronic temperature T$_e (\tau)$.
Figure\,\ref{Fig4}(a1) shows the nodal differential map for Bi2201-OD24, [$\int_k \text{I}(\omega,\textbf{k},\tau >0)dk - \int_k \text{I}(\omega,\textbf{k},\tau <0)dk$]. From this map, the population and depletion dynamics can be extracted independently, as shown by open circles in Fig.\,\ref{Fig4}(c) (integration window of 50\,meV above and below E$_\mathrm{F}$, respectively). We note that, although population and depletion share comparable decay constants ($\sim$0.85\,ps and $\sim$4.5\,ps, as obtained via a phenomenological double exponential fit for both population and depletion), their amplitudes differ by a factor two. Such a difference stems from the combined contributions of $A(\omega,\tau)$ and $f(\omega,\tau)$ to $\int_k \text{I}(\omega,\textbf{k},\tau)dk$, and does not imply any particle-hole asymmetry. 
Within the FL description, an increase in the electronic temperature T$_e (\tau)$ broadens $A(\omega,\tau)$, and suppresses the CSW equally above and below E$_\mathrm{F}$. The carrier redistribution encoded in $f(\omega,\tau)$ is asymmetric, leading to an increased photoemission intensity above E$_\mathrm{F}$, instead of a reduction below. Together, these two effects are what result effectively in the factor of two difference between population and depletion signals.

Using the transient electronic temperature T$_e (\tau)$ of Fig.\,\ref{Fig4}(b) within the Fermi liquid framework of Eq.\,\ref{eq1FL}, we compute the simulated analog to the differential momentum-integrated energy vs. delay map [Fig.\,\ref{Fig4}(a2)], as well as the population/depletion traces [Fig.\,\ref{Fig4}(c), dashed lines]. The quantitative agreement between experimental and simulated population/depletion dynamics demonstrates that a description in terms of a transient thermally-broadened state defined by T$_e (\tau)$ and the FL coefficients $\alpha$ and $\beta$ (see Fig.\,\ref{Fig3}) reproduces the relaxation timescales, and also captures the different amplitudes of the population and depletion signals.
This agreement further confirms the fundamental role played by the normal state electrodynamics in defining underlying electronic correlations along the gapless nodal direction in superconducting cuprates. Moreover, it highlights the different contributions to the transient photoemission intensity \inserted{already glimpsed in the intensity redistribution across the Brillouin Zone shown in Fig.\,\ref{Fig2}(d)}. The latter cannot be described uniquely in terms of population/depletion dynamics controlled by the effective Fermi-Dirac distribution function $f(\omega,\tau)$; instead, one should take into account the spectral function contributions to the photoemission intensity, which are always present in all TR-ARPES experiments.

\section{Photoinduced filling of the superconducting gap}
After having explored the suppression of coherent spectral weight and the related transient evolution of the photoemission intensity along the nodal direction of single- and bi-layer Bi-based cuprates, we now move on to the investigation of the pump-induced evolution of the superconducting gap, as shown in Fig.\,\ref{Fig5}(a).
We have recently demonstrated the filling of the superconducting gap in the near-nodal region of Bi2212-UD82 via photo-melting of the coherence of the macroscopic condensate, due to non-thermal bosons \cite{boschini2018}. These results have offered compelling evidence for the primary role of phase coherence in the formation of the superconducting condensate in cuprates, while the thermal occupation of above-gap states plays a secondary role \cite{emery_kivelson_1995,reber2012TDOS,kondo2015NatComm}. 

We first provide additional experimental evidence of the non-thermal nature of the filling of the superconducting gap by performing the same analysis of Ref.\,\onlinecite{boschini2018,kondo2015NatComm} on Bi2212-OP91. In particular, we focus our investigation on the transient evolution of the superconducting gap approximately 14$^o$ off the nodal direction ($\varphi$=31$^o$), just before the endpoint of the Fermi arc \cite{vishik2012PNAS}. Figure\,\ref{Fig5}(a) displays the ARPES mapping at $\varphi$=31$^o$ for negative delays ($\tau <$0\,ps, panel c), and at the maximum gap filling time ($\tau$=0.6\,ps, panel d). 
The filling of the superconducting gap at $\mathbf{k}_\mathrm{F}$ is captured well by the phenomenological self-energy proposed by Norman et al. \cite{norman1998SelfEnergy}:
\begin{equation}
    \label{EQ2}
    \Sigma=-i\Gamma_s + \frac{\Delta^2}{\omega+ i \Gamma_p},
\end{equation}
where $\Gamma_s$ is the single-particle scattering rate, $\Delta$ the superconducting gap, and $\Gamma_p$ the pair breaking scattering term. Note that although the phenomenological self-energy of Eq.\,\ref{EQ2} violates the sum rule for the occupation number \cite{herman2017PRB_theoryExtended}, it can be considered a reasonable approximation at $\mathbf{k}_\mathrm{F}$ \cite{kondo2015NatComm,boschini2018}.

In an effort to highlight the transient filling of the superconducting gap by simple visual inspection, we apply the tomographic density of states method (TDOS) proposed by Reber et al. \cite{reber2012TDOS} into the time domain.
The TDOS is computed by normalizing the momentum-integrated off-nodal intensity by the nodal counterpart, and it is directly proportional to the Dynes function \cite{dynes1978_PRL,reber2012TDOS}.
The time-dependent TDOS presented in Fig.\,\ref{Fig5}(b) shows that the spectral weight is indeed transferred from the coherent peaks into the gap region, while the gap amplitude is only marginally affected. However, one should note that the Dynes function is defined by only two terms, namely the gap amplitude $\Delta$ and a scattering term $\Gamma_{\text{Dynes}}$, and this latter can be expressed as $\Gamma_{\text{Dynes}}=(\Gamma_s+\Gamma_p)/2$ when using the self-energy of Eq.\,\ref{EQ2} \cite{herman2017PRB_theoryExtended}. This implies that, despite providing a direct visualization of the transient filling of the superconducting gap, the TDOS method cannot uniquely disentangle the contributions of single-particle and pair-breaking scattering terms and, consequently, determine whether either (or both) of them have a non-thermal character \cite{parham2017prx}.

With the goal of revealing the transient dynamics of all terms of Eq.\,\ref{EQ2}, we display in Fig.\,\ref{Fig5}(c) off-nodal symmetrized EDCs at the Fermi momentum $\mathbf{k}_\mathrm{F}$ for various pump-probe delays (note that the EDCs have been deconvoluted from the energy resolution using the Lucy-Richardson algorithm \cite{yang2008NatureDeconv,razzoli2013PRL}). 
The self-energy of Eq.\,\ref{EQ2} reproduces well the progression of the symmetrized EDCs upon pump excitation [fits shown in Fig.\,\ref{Fig5}(c) by solid lines], allowing us to extract the transient evolution of each individual parameter in Eq.\,\ref{EQ2}. Figure\,\ref{Fig5}(d) displays the normalized variation of $\Gamma_s$, $\Delta$, and $\Gamma_p$ (left), alongside the electronic temperature as extracted along the nodal direction (filled black circles, right axis) as a function of the pump-probe delay.
While the gap amplitude and the single-particle scattering rate mimic the temporal evolution of the electronic temperature, thus establishing their evolution as thermally driven, the pair-breaking scattering term exhibits remarkably different dynamics. In agreement with what was previously reported in Ref.\,\onlinecite{boschini2018}, $\Gamma_p$ peaks at $\sim$0.5\,ps after the zero pump-probe delay, and is defined by a decay constant of $\sim$1\,ps. The dichotomy between T$_e (\tau)$ and $\Gamma_p (\tau)$ demonstrates the non-thermal origin of the observed pump-induced melting of the coherent condensate. 

\begin{figure}
\centering
\includegraphics[scale=1]{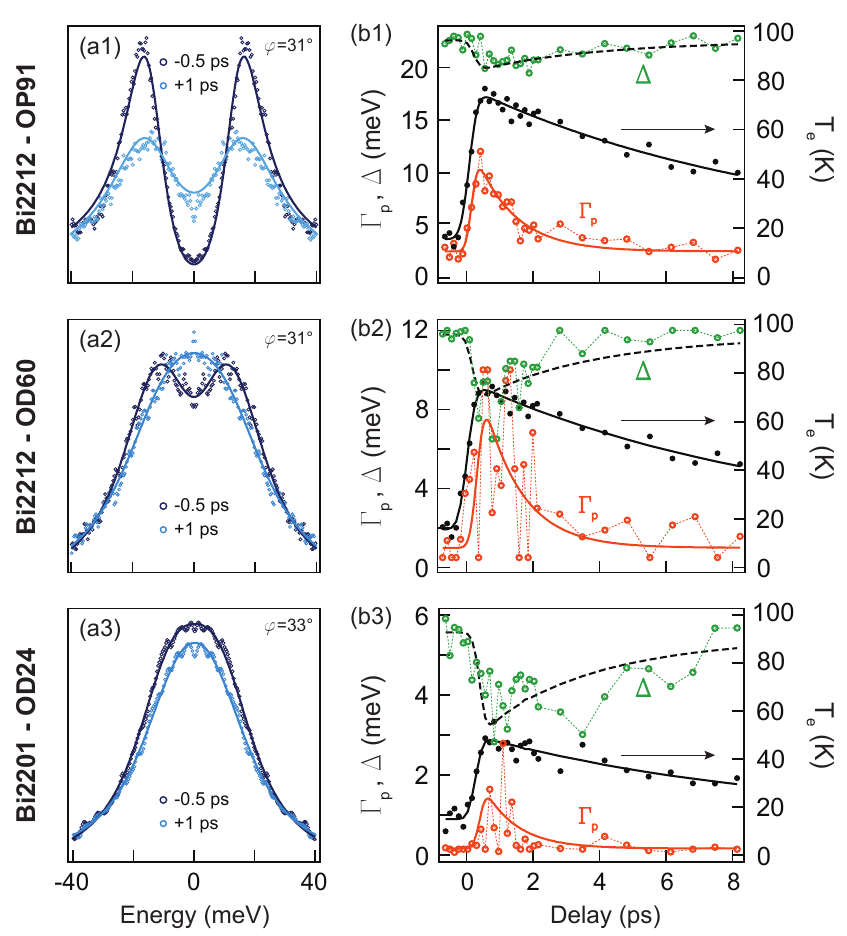}
\caption[Fig2]{(a1) Symmetrized EDCs of Bi2212-OP91 along the off-nodal direction $\varphi$=31$^o$. Solid lines are fits via the extended self-energy of Eq.\,\ref{EQ3}-\ref{EQ4}, using the electronic temperature and FL parameters $\alpha$ and $\beta$ obtained along the nodal direction (see Fig.\,\ref{Fig3}). (b1) Transient evolution of $\Delta$ and $\Gamma_p$ in Eq.\,\ref{EQ3}-\ref{EQ4} extracted via fitting of symmetrized EDCs (left axis), and T$_e (\tau)$ used in the fit, for Bi2212-OP91 (right axis). Orange and black solid lines are phenomenological double-exponential decay fits to the data; the black dashed line represents the expected evolution of a BCS-like gap assuming a closure of the gap at 30$\%$ above T$_c$. (a2)-(b2) Same as (a1)-(b1) but for overdoped Bi2212 (T$_c$=$60\,$K, OD60). (a3)-(b3) Same as (a1) and (b1) but for Bi2201-OD24 along $\varphi$=33$^o$ from the nodal direction.}
\label{Fig6}
\end{figure}

Before proceeding to the detailed discussion of the possible mechanisms underlying the transient enhancement of $\Gamma_p (\tau)$, 
we first point out that one of the main shortcomings of the phenomenological self-energy in Eq.\,\ref{EQ2} is the assumption of $\omega$-independent scattering rates. However, based on our previous analysis of the nodal coherent spectral weight (Figs.\,\ref{Fig3}-\ref{Fig4}), and on the close similarity in the temporal evolution of $\Gamma_s$ and T$_e$ (Fig.\,\ref{Fig5}), we can infer that the low-energy electrodynamics is well described by the Fermi liquid framework in the near-nodal region  \cite{chang2013NatComm_anisotropic}. Accordingly, we propose an extended self-energy ($\Sigma_\text{ext}$) by combining Eqs.\,\ref{eq1FL} and\,\ref{EQ2}, which is defined as follows: 
\begin{equation}
    \label{EQ3}
    \Sigma'_{\text{ext}}(\omega,\text{T}_e)=\Sigma'_{\text{FL}}(\omega,\text{T}_e) + \frac{\omega \Delta^2 }{\omega^2+\Gamma_p^2},
\end{equation}
\begin{equation}
    \label{EQ4}
    \Sigma''_{\text{ext}}(\omega,\text{T}_e)= \Sigma''_{\text{FL}}(\omega,\text{T}_e) -\frac{\Delta^2 \Gamma_p}{\omega^2+\Gamma_p^2}. 
\end{equation}

This phenomenological self-energy mitigates the problems of Eq.\,\ref{EQ2} by incorporating the $\omega$- and T-dependence of the single-particle scattering rate via FL theory (note that $\Sigma_{\text{ext}}$ satisfy causality since both Eqs.\,\ref{eq1FL} and\,\ref{EQ2} themselves comply with Kramers–Kronig relations). By using T$_e (\tau)$ and the FL parameters ($\alpha$, $\beta$, $\Gamma_{\text{imp}}$) extracted along the nodal direction, we employ Eqs.\,\ref{EQ3},\ref{EQ4} for tracking the temporal evolution of $\Delta$ and $\Gamma_p$ in Bi2212 OP91 and OD60, as well as in Bi2201-OD24 (Fig.\,\ref{Fig6}). Not only does this extended phenomenological approach grant stability in the fitting procedure, even for gap amplitudes well below our energy resolution (17\,meV), but it confirms the non-thermal origin of $\Gamma_p$, which shows similar dynamics for the three dopings presented in Fig.\,\ref{Fig6}. Moreover, the black dashed lines in Fig.\,\ref{Fig6}(b),(d),(f) display the predicted thermal suppression of the gap according to the conventional BCS gap equation and assuming the closure of the gap at $\sim$30$\%$ above T$_c$, in agreement with Ref.\,\onlinecite{kondo2015NatComm,reber2012TDOS}.

In summary, our TR-ARPES study of Bi-based cuprates demonstrates that while the transient evolution of the gap and single-particle scattering rate stem from a mere thermal effect, the pump-induced enhancement of $\Gamma_p (\tau)$ has a non-thermal origin and is independent of the doping level and number of CuO$_2$ layers. The $\sim$500\,fs delay in the peak of $\Gamma_p (\tau)$ with respect to the pump excitation can be explained in terms of the time needed for the development of a non-thermal bosonic population. In a nutshell, the ultrafast relaxation of photoexcited non-thermal quasiparticles is accompanied by the emission of high-energy bosons \cite{perfetti2007,rameau2016NatComm,dalConte2012Science}, whose re-absorption may break additional Cooper pairs and transiently increase $\Gamma_p$. In parallel, any pair-recombination process emits a boson. The interplay between bosonic absorption/emission processes leads to a delayed maximum of the non-thermal bosonic population at $\sim$500\,fs, a feature which is well captured by the Rothwarf–Taylor equations \cite{giannetti2016review,boschini2018}.
However, we note that the $\sim$1\,ps recovery time of $\Gamma_p (\tau)$ does not show any clear doping- and layer-dependence, and matches the phase-correlation time extracted from high-frequency conductivity \cite{corson1999Nature}. We may speculate that the $\sim$1\,ps relaxation time of $\Gamma_p$ represents a universal timescale for pair-breaking bosons in Bi-based cuprates, although further studies of the non-thermal evolution of the superconducting gap and corresponding phase coherence are needed to verify this speculation. A further extension of Eqs.\,\ref{EQ3},\ref{EQ4} by including the energy dependence of $\Gamma_p$, as well as complementary time-resolved x-ray scattering \cite{gerber2017Science,wandel2020arXiv,mitrano2020CommPhys} and electron diffuse scattering \cite{CotretPRB2019_UEDSgraphite} experiments, may reveal which bosons, if any, are transiently coupled to the superconducting condensate.

\section{Conclusion and outlook}
In conclusion, although a comprehensive understanding of the TR-ARPES signal may be challenging, extensive efforts in the past two decades have demonstrated the exquisite power of the TR-ARPES technique in the study of quantum materials. Here we focused our attention on copper oxides high-temperature superconductors, which represent a prototypical platform for the exploration of strong-correlation many-body phenomena and novel quantum phases of matter.  
We have briefly reviewed how TR-ARPES can access valuable information regarding quasiparticles, the superconducting gap, and band-dispersion dynamics, as well as many other aspects of the cuprate phase diagram.
Thereafter, we discussed our recent work on the temperature-driven suppression of the nodal coherent spectral weight in Bi-based cuprates in terms of the Fermi liquid framework \cite{zonno2021PRB}. In particular, we demonstrated that the observed nodal TR-ARPES signal encompasses two main contributions, namely the quasiparticle dynamics (encoded in the transient electronic distribution), as well as the pump-induced self-energy modifications (manifesting as spectral weight redistribution and broadening). We confirmed the non-thermal nature of the pump-induced gap filling in the Bi-based cuprates family, in agreement with Ref.\,\onlinecite{boschini2018}.
Specifically, while we established that gap dynamics and amplitude are locked to the evolution of the electronic occupation, \text{i.e.} the electronic temperature, the transient filling of the superconducting gap reflects pair-breaking phenomena driven by non-thermal bosons that quench the phase coherence of the macroscopic superconducting condensate. In addition, we reported similar gap-filling dynamics for different doping levels and compounds (single- and bi-layer alike), pointing towards a universal mechanism for the superconducting-to-normal state phase transition in Bi-based cuprates.

As a final outlook, we remark that although most of the TR-ARPES studies carried out so far on cuprates have relied on near $\sim$\,6\,eV probe systems with near-IR pump excitation, in the coming years we foresee new and fascinating explorations of ultrafast dynamics and light-induced phases of matter.
The development of high-repetition rate and high-energy/temporal resolution TR-ARPES systems with high-harmonic XUV probes \cite{sie2019NatComm,mills2019RSI,lee2020RSI,puppin2019RSI}, and pump excitations in the mid-IR/THz range \cite{wang2013Science,reimann2018subcycle,aeschlimann2021GierzArXiv}, will surely pave the way towards exciting and unprecedented investigations of out-of-equilibrium properties of quantum materials over the full momentum space. The most exciting times lie ahead!

\begin{acknowledgments}
We gratefully acknowledge S. K. Y. Dufresne, M. X. Na, M. Bluschke, E. Razzoli, M. Michiardi, A. K. Mills, S. Zhdanovich, G. Levy, C. Giannetti and D. J. Jones for insightful discussions, reviewing the manuscript, and their contributions in acquiring, analyzing, and interpreting the data. We are also grateful to Y. Yoshida, H. Eisaki (Bi2201), and G. D. Gu (Bi2212) for providing high-quality single-crystals.
This research was undertaken thanks in part to funding from the Max Planck-UBC-UTokyo Centre for Quantum Materials and the Canada First Research Excellence Fund, Quantum Materials and Future Technologies Program. This project is also funded by the Gordon and Betty Moore Foundation's EPiQS Initiative, Grant GBMF4779 to A.D. and D.J.J.; the Killam, Alfred P. Sloan, and Natural Sciences and Engineering Research Council of Canada's (NSERC's) Steacie Memorial Fellowships (A.D.); the Alexander von Humboldt Fellowship (A.D.); the Canada Research Chairs Program (A.D.); NSERC, Canada Foundation for Innovation (CFI); British Columbia Knowledge Development Fund (BCKDF); and the CIFAR Quantum Materials Program.
\end{acknowledgments}

\bibliographystyle{apsrev4-1}

\providecommand{\noopsort}[1]{}\providecommand{\singleletter}[1]{#1}

\end{document}